# Nonadditive Conditional Entropy and Its Significance for Local Realism

Sumiyoshi Abe (1) and A. K. Rajagopal (2)

(1)*College of Science and Technology*, *Nihon University*,

*Funabashi*, *Chiba 274-8501*, *Japan*

(2)*Naval Research Laboratory*, *Washington*, *D. C. 20375-5320*, *USA*

Based on the form invariance of the structures given by Khinchin's axiomatic foundations of information theory and the pseudoadditivity of the Tsallis entropy indexed by $q$, the concept of conditional entropy is generalized to the case of nonadditive (nonextensive) composite systems. The proposed nonadditive conditional entropy is classically nonnegative but can be negative in the quantum context, indicating its utility for characterizing quantum entanglement. A criterion deduced from it for separability of density matrices for validity of local realism is examined in detail by employing a bipartite spin-1/2 system. It is found that the strongest criterion is obtained in the limit $q \to \infty$.

In this Letter, we present a generalization of the concept of "conditional entropy" for nonadditive (nonextensive) composite systems. This is done based on the form invariance of the structures in Khinchin's axiomatic foundations of classical information theory and the pseudoadditivity of the Tsallis entropy indexed by $q$ in. nonextensive statistical mechanics. Then, we discuss the nonadditive conditional entropy in the quantum context. It is shown that the nonadditive conditional entropy is nonnegative for all classically correlated states (i.e., separable states) but can be negative for nonclassical correlated states (i.e., quantum entangled states). Since classically correlated states admit local hidden-variable models, the positivity of the nonadditive conditional entropy leads to a criterion for validity of local realism. This criterion depends on the Tsallis nonadditivity parameter $q$. We shall see how the strongest criterion can be obtained by controlling the value of $q$. The present attempt can also be thought of as a stepping stone towards nonadditive (quantum) information theory.

In the development of the foundations of classical information theory, Khinchin [1] presented a mathematically rigorous proof of a uniqueness theorem for the Boltzmann-Shannon entropy based on the additivity law for a composite system in terms of the concept of conditional entropy. Suppose the total system be divided into two subsystems, $A$ and $B$, and let $p_{ij}(A, B)$ be the normalized joint probability of finding $A$ and $B$ in their $i$th and $j$th states, respectively. Then the conditional probability of $B$ with $A$ found in its $i$th state is given by $p_{ji}(B|A) = p_{ij}(A, B) / p_i(A)$, which leads to the celebrated Bayes multiplication law

$$p_{ij}(A, B) = p_i(A) p_{ij}(B|A), \tag{1}$$

where $p_i(A)$ is the marginal probability distribution: $p_i(A) = \sum_j p_{ij}(A, B)$. It should be noted that this form of factorization can always be established in any physical situation. From this law, the Boltzmann-Shannon entropy

$$S(A, B) = -\sum_{i,j} p_{ij}(A, B) \ln p_{ij}(A, B) \tag{2}$$

$$S(A, B) = S(A) + S(B|A), \tag{3}$$

where $S(B|A)$ stands for the conditional entropy defined by

$$S(B|A) = \langle S(B|A_i) \rangle^{(A)} = \sum_i p_i(A) S(B|A_i) \tag{4}$$

with

$$S(B|A_i) \equiv -\sum_j p_{ij}(B|A) \ln p_{ij}(B|A). \tag{5}$$

In the particular case when $A$ and $B$ are statistically independent, $p_{ij}(B|A)$ is equal to $p_j(B)$ and therefore $S(B|A) = S(B)$, implying the additivity law: $S(A, B) = S(A) + S(B)$. We emphasise here that there is *a natural correspondence relation between the multiplication law and the additivity law*:

$$p_{ij}(A, B) = p_i(A) p_{ij}(B|A) \leftrightarrow S(A, B) = S(A) + S(B|A). \tag{6}$$

When the above discussion is generalized to quantum theory of composite systems, a remarkable feature appears. The conditional entropy may then take negative values, suggesting its importance for characterizing quantum entanglement [2]. This feature may have its root in a profound difference between classical and quantum probability concepts: in the latter, one of Kolmogorov's axioms, namely the additivity of the probability measure, is violated in general [3]. Also, there are theoretical observations [4,5] that formal correspondences exist between thermodynamics and quantum entanglement. These investigations seem to suggest that *the measure of quantum entanglement may not be additive* [5,6].

Nonadditivity, or nonextensivity, is an important concept also in the field of statistical mechanics. A statistical system is nonextensive if it contains long-range interaction, long-range memory, or (multi)fractal structure. In such a system, a macroscopic thermodynamic quantity (e.g., the internal energy) is not simply proportional to the microscopic degrees of freedom. Boltzmann-Gibbs statistical mechanics, which can be constructed based on the additive Boltzmann-Shannon entropy is known to expose difficulties, if applied to such a system [7]. In this respect, a nonextensive generalization of Boltzmann-Gibbs statistical mechanics formulated by

Tsallis [8,9] has attracted much attention in recent years. In this formalism, referred to as nonextensive statistical mechanics, the Boltzmann-Shannon entropy in eq. (2) is generalized as follows:

$$S_q(A, B) = \frac{1}{1-q}\left\{\sum_{i,j}\left[p_{ij}(A, B)\right]^q - 1\right\}, \tag{7}$$

where $q$ is a positive parameter. This quantity converges to the Boltzmann-Shannon entropy in the limit $q \to 1$. Like the Boltzmann-Shannon entropy, it is nonnegative, possesses the definite concavity for all $q > 0$, and is known to satisfy the generalized H-theorem. Nonextensive statistical mechanics has found a lot of physical applications. (A comprehensive list of references is currently available at URL [10]. See also Ref. [11].)

A standard discussion about the nonadditivity of the Tsallis entropy $S_q[p]$ assumes factorization of the joint probability distribution in eq. (2). Then, the Tsallis entropy is found to yield the so-called pseudoadditivity relation

$$S_q(A, B) = S_q(A) + S_q(B) + (1-q)S_q(A)S_q(B). \tag{8}$$

Clearly, the additivity holds if and only if $q \to 1$. However, there is a logical difficulty in this discussion. As mentioned above, Tsallis' nonextensive statistical mechanics was devised in order to treat a statistical system with, for example, a long-range interaction. On the other hand, physically, "dividing the total system into the subsystems" implies that the subsystems are made spatially separated in such a way that there is no residual interaction or correlation. If the system is governed by a long-range interaction, the statistical independence can never be realized by any spatial separation since the influence of the interaction persists at all distances. In fact, the probability distribution in nonextensive statistical mechanics does not have a factorizable form even if the systems $A$ and $B$ are dynamically independent, and therefore correlation is always induced by nonadditivity of statistics [12]. Another important physical mechanism which prevents realizing the statistical independence by spatial separation is provided by the concept of entanglement in quantum theory. Thus, it is clear that the assumption of the factorized joint probability distribution is not physically pertinent for characterizing the nonadditivity of the Tsallis entropy. These considerations naturally lead us to the necessity of defining the conditional entropy associated with the Tsallis entropy.

To overcome the above-mentioned logical difficulty and to generalize the correspondence relation in eq. (6) simultaneously, first let us recall the definition of the normalized $q$-expectation value in nonextensive statistical mechanics [9]. In order to specify the constraint on a physical quantity to develop the Jaynes maximum entropy principle for nonextensive statistical mechanics based on the Tsallis entropy, the authors of Ref. [9] introduced the idea of the normalized $q$-expectation value. Consider a physical quantity $Q$, whose value in the system's $i$th state is denoted by $Q_i$. Its normalized $q$-expectation value is defined by

$$<Q>_q = \sum_i P_i Q_i, \qquad (9)$$

where $P_i$ is the so-called escort distribution [13] associated with the original probability distribution $p_i$:

$$P_i = \frac{(p_i)^q}{\sum_i (p_i)^q}. \qquad (10)$$

Next, let us consider the Tsallis entropy of the conditional probability distribution

$$S_q(B|A_i) = \frac{1}{1-q}\left[\sum_j \left(p_{i,j}(B|A)\right)^q - 1\right]$$

$$= \frac{1}{1-q}\left[\frac{\sum_j \left(p_{i,j}(A,B)\right)^q}{\left(p_i(A)\right)^q} - 1\right], \qquad (11)$$

which is a natural nonadditive generalization of eq. (5). Now, in conformity with eq. (4), we propose to define the nonadditive conditional entropy as the normalized $q$-expectation value of $S_q(B|A_i)$ with respect to the marginal probability distribution $p_i(A)$ as follows:

$$S_q(B|A) \equiv \left\langle S_q(B|A_i) \right\rangle_q^{(A)}$$

$$= \frac{\sum_i (p_i(A))^q S_q(B|A_i)}{\sum_i (p_i(A))^q}. \tag{12}$$

Using eq. (11) in eq. (12), we obtain

$$S_q(B|A) = \frac{S_q(A, B) - S_q(A)}{1 + (1-q)S_q(A)}. \tag{13}$$

This is our definition of the nonadditive conditional entropy. From this, it is immediate to see

$$S_q(A, B) = S_q(A) + S_q(B|A) + (1-q)S_q(A)S_q(B|A), \tag{14}$$

which is a natural nonadditive generalization of eq. (3) in view of the pseudoadditivity in eq. (8). Therefore, the correspondence relation in eq. (6) is generalized to

$$p_{ij}(A, B) = p_i(A)p_{ij}(B|A) \leftrightarrow$$
$$S_q(A, B) = S_q(A) + S_q(B|A) + (1-q)S_q(A)S_q(B|A). \tag{15}$$

In quantum theory, the probability distribution is replaced by the density matrix $\hat{\rho}$, which is Hermitian, traceclass, and positive semidefinite. It incorporates both pure and mixed states of a system. When the state is pure, the equality $\hat{\rho}^2 = \hat{\rho}$ holds, whereas $\hat{\rho}^2 < \hat{\rho}$ for the mixed state. The quantum counterpart of eq. (7) is given by

$$S_q(A, B) = \frac{1}{1-q}\left[\text{Tr}(\hat{\rho}(A, B))^q - 1\right], \tag{16}$$

which converges to the von Neumann entropy $S(A, B) = -\text{Tr}[\hat{\rho}(A, B) \ln \hat{\rho}(A, B)]$ in the limit $q \to 1$. It is evident that, like the von Neumann entropy, this quantity is nonnegative and vanishes for a pure state.

There is an ordering ambiguity in defining the "conditional density matrix", since the joint density matrix and its marginals do not commute with each other, in general. Here, we avoid to define it explicitly and directly translate eq. (13) to its quantum counterpart. That is,

$$S_q(B|A) = \frac{S_q(A, B) - S_q(A)}{1 + (1-q)S_q(A)}, \tag{17}$$

with

$$S_q(A) = \frac{1}{1-q}\left[\text{Tr}_A(\hat{\rho}(A))^q - 1\right], \tag{18}$$

and $S_q(A, B)$ given in eq. (16). In eq. (18), $\hat{\rho}(A)$ is the marginal density operator defined by $\hat{\rho}(A) = \text{Tr}_B \hat{\rho}(A, B)$, where $\text{Tr}_A$ ($\text{Tr}_B$) stands for the partial trace over the states of the subsystem $A$ ($B$). By going to the representations in which $\hat{\rho}(A, B)$ and $\hat{\rho}(A)$ are diagonal, $S_q(B|A)$ appears to take the form of eq. (12). The difference between the two is now that the expansion coefficients are the eigenvalues of the diagonalized density matrices in contrast to their classical counterparts presented earlier. In the limit $q \to 1$, $S_q(B|A)$ converges to the conditional von Neumann entropy $S(B|A) = S(A, B) - S(A)$ discussed in detail in Ref. [2].

Let us consider two particular cases: a product state and a classically correlated state. A product state $\hat{\rho}(A, B) = \hat{\rho}(A) \otimes \hat{\rho}(B)$, gives rise to the pseudoadditivity relation, whose form is the same as the classical one in eq. (8). Therefore, we have $S_q(B|A) = S_q(B)$ as expected. A classically correlated state, i.e. a separable state, is a convex combination of product states [14]:

$$\hat{\rho}(A, B) = \sum_\lambda w_\lambda \, \hat{\rho}_\lambda(A) \otimes \hat{\rho}_\lambda(B), \tag{19}$$

where $w_\lambda \in [0, 1]$ and $\sum_\lambda w_\lambda = 1$. This state is known to admit local hidden-variable models and to satisfy the Bell inequality [14]. Using the orthonormal bases of $A$ and $B$, we write

$$\hat{\rho}_\lambda(A) = \sum_a p_\lambda(a)|a\rangle\langle a|, \qquad \hat{\rho}_\lambda(B) = \sum_b r_\lambda(b)|b\rangle\langle b|, \tag{20}$$

where $p_\lambda(a), r_\lambda(b) \in [0, 1]$ and $\sum_a p_\lambda(a) = \sum_b r_\lambda(b) = 1$. In this case, the nonadditive quantum conditional entropy is given by

$$S_q(B|A) = \frac{\sum_a \left[\sum_\lambda w_\lambda p_\lambda(a)\right]^q S_q(B|a)}{\sum_a \left[\sum_\lambda w_\lambda p_\lambda(a)\right]^q}, \tag{21}$$

provided that we have used the notation $S_q(B|a) = (1-q)^{-1}\left\{\sum_b \left[\pi(b|a)\right]^q - 1\right\}$ with $\pi(b|a) \equiv \sum_\lambda w_\lambda p_\lambda(a) r_\lambda(b) / \sum_\lambda w_\lambda p_\lambda(a)$. Note that $\pi(b|a) \in [0, 1]$ and $\sum_b \pi(b|a) = 1$.

Therefore $\pi(b|a)$ is analogous to the classical conditional probability distribution. This fact makes $S_q(B|a)$ nonnegative. Thus, we conclude that $S_q(B|A) \geq 0$ for any classically correlated states. This establishes a $q$-generalization of the result in Ref. [2].

Here, it is worth mentioning that in Ref. [15] the Rényi entropy $S_\alpha^R[\hat{\rho}] = (1-\alpha)^{-1} \times \ln \text{Tr}(\hat{\rho})^\alpha$ with $\alpha > 1$ is discussed in connection with violation of local realism in quantum theory. The authors of Ref. [15] proposed the $\alpha$-entropic inequality: $S_\alpha^R(A, B) \geq \max\{S_\alpha^R(A), S_\alpha^R(B)\}$, which can be violated by entangled mixed states. We note that the Rényi entropy is additive but is not concave for $\alpha > 1$.

A corresponding entropic inequality in the present nonadditive conditional entropy is

$$S_q(B|A) \geq 0, \qquad S_q(A|B) \geq 0. \tag{22}$$

As mentioned above, eq. (22) holds for all classically correlated states that can be modelled by local hidden-variable theories. Nonclassical correlated states, that is, quantum entangled states, yield $S_q(B|A) < 0$. Therefore, the inequality (22) can be used as a criterion for separability of density matrices. A point is that the inequality depends on the parameter $q$.

To see how the separability criterion is strengthened by controlling the value of $q$, let us employ a bipartite spin-1/2 system. For this purpose, we consider the following parametrized form of the Werner-Popescu state [14,15]:

$$\hat{\rho}(A, B) = \frac{1-x}{4} \hat{I}_A \otimes \hat{I}_B + x |\Psi^-\rangle\langle\Psi^-|, \tag{23}$$

where $0 \leq x \leq 1$, $\hat{I}_A$ ($\hat{I}_B$) the $2 \times 2$ identity matrix in the space of the spin $A$ ($B$), and

$$|\Psi^-\rangle = \frac{1}{\sqrt{2}} \left( |\uparrow\rangle_A |\downarrow\rangle_B - |\downarrow\rangle_A |\uparrow\rangle_B \right). \tag{24}$$

The separability criteria known so far are: a) $x < 1/\sqrt{2}$ (the Bell inequality), b) $x < 1/\sqrt{3}$ (the $\alpha$-entropic inequality with $\alpha = 2$) [16], and c) $x < 1/3$ (the nonnegative eigenvalues of the density matrix with partial transpose) [17]. The strongest criterion c) found by Peres [17] is known to actually be the necessary and sufficient condition for the separability of the bipartite spin-1/2 system [18]. Our interest here is to see whether the Peres criterion obtained by an algebraic method can be derived within the present framework of nonadditive quantum information theory. From the density matrix in eq. (23), the nonadditive conditional entropy is calculated to be

$$S_q(B|A) = S_q(A|B) = \frac{1}{1-q}\left[\frac{3}{2}\left(\frac{1-x}{2}\right)^q + \frac{1}{2}\left(\frac{1+3x}{2}\right)^q - 1\right]. \tag{25}$$

For a fixed value of $q$, this is a monotonically decreasing function of $x$. In Fig. 1, we present an implicit plot of $S_q(B|A) = 0$ with respect to $q \in [0, \infty)$ and $x \in [0, 1]$. One clearly sees how the criterion obtained from the conditional von Neumann entropy ($q \to 1$) can be strengthened by increasing the value of $q$. From eq. (25), it is also evident that $S_q(B|A) \to 0$ in the limit $q \to \infty$ if and only if $x < 1/3$. Thus, the Peres strongest criterion is obtained from the present information-theoretic approach.

In conclusion, we have constructed the nonadditive conditional entropy based on the form invariance of the structures of the axiomatic foundations of classical information theory and the pseudoadditivity of the Tsallis entropy indexed by $q$. Then, we have discussed it in the quantum context and applied it to separability of the density matrix for validity of local realism. We have found that for a bipartite spin-1/2 system, i.e., a $2 \times 2$ system, the positivity of the nonadditive conditional entropy leads to the Peres strongest criterion for separability in the limit $q \to \infty$. In Ref. [18], it has been discussed that Peres's method of partial transposition of the density matrix yields the necessary and sufficient condition for separability of $2 \times 2$ and $2 \times 3$ systems but not for other general systems. Therefore, for a further development, it is important to clarify the general properties of the condition in eq. (22).

S.A. was supported by the GAKUJUTSU-SHO Program of College of Science and Technology, Nihon University. A.K.R. acknowledges the partial support of the U.S. Office of Naval Research.

# Figure Caption

An implicit plot of $S_q(B|A) = 0$ with respect to $q \in [0, \infty)$ and $x \in [0, 1]$. In the limit $q \to \infty$, $x$ converges to $1/3$.

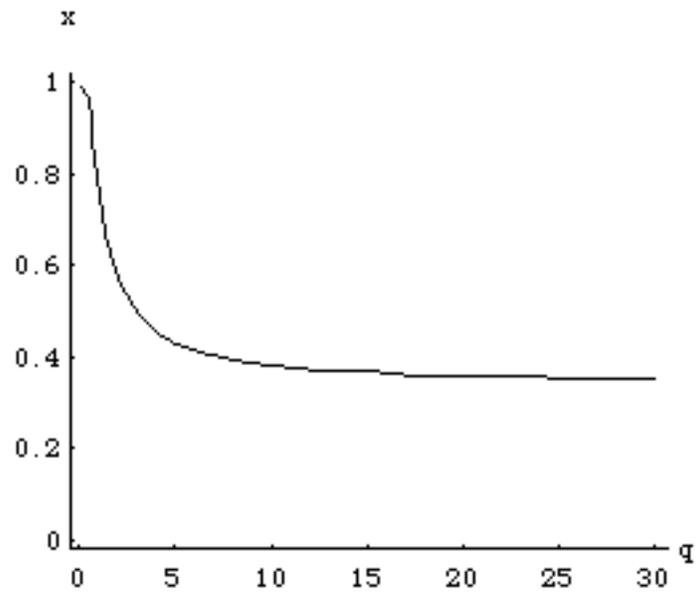

Fig. 1